\documentclass[12pt]{article}

\usepackage{scicite}
\newcommand\ddfrac[2]{\frac{\displaystyle #1}{\displaystyle #2}}

\usepackage{dirtytalk}
\usepackage{graphicx}
\graphicspath{ {images/} }
\usepackage{float}
\usepackage{times}
\usepackage{lineno}
\usepackage[normalem]{ulem}
\usepackage{xcolor}

\usepackage{hyperref}
\usepackage{amsfonts} 

\usepackage{changepage}
\usepackage{amsmath}
\usepackage{array}
\usepackage[super]{nth}

\def\epsilon{\varepsilon}

\newenvironment{sciabstract}{%
\begin{quote} \bf}
{\end{quote}}
\title{Node-based Generalized Friendship Paradox fails}

\author
{Anna Evtushenko,$^{1\ast}$ Jon Kleinberg$^{1, 2}$\\
\\
\normalsize{$^{1}$Department of Information Science, Cornell University, Ithaca NY, USA}\\
\normalsize{$^{2}$Department of Computer Science, Cornell University, Ithaca NY, USA}\\
\\
\normalsize{$^\ast$anna@infosci.cornell.edu}
}

\date{}

\begin{document}

\baselineskip24pt

\maketitle

\begin{sciabstract}
The Friendship Paradox---the principle that \say{your friends have more friends than you do}---is a combinatorial fact about degrees in a graph; but given that many web-based social activities are correlated with a user's degree, this fact has been taken more broadly to suggest the empirical principle that \say{your friends are also more active than you are.} This Generalized Friendship Paradox, the notion that any attribute positively correlated with degree obeys the Friendship Paradox, has been established mathematically in a network-level version that essentially aggregates uniformly over all the edges of a network.

Here we show, however, that the natural node-based version of the Generalized Friendship Paradox---which aggregates over nodes, not edges---may fail, even for degree-attribute correlations approaching 1. Whether this version holds depends not only on degree-attribute correlations, but also on the underlying network structure and thus can't be said to be a universal phenomenon. We establish both positive and negative results for this node-based version of the Generalized Friendship Paradox and consider its implications for social-network data.
\end{sciabstract}




\pagestyle{plain}

\section{Introduction}
\label{sec:intro}
The Friendship Paradox is a well-known graph-theoretical statement
about the relationship between nodes' degrees (their number of neighbors or friends, a first-order quantity)
and their friends' degrees (a second-order quantity) in a simple
connected finite graph \cite{feld1991your}. In fact, in any such
graph, the mean of the latter is no smaller than the mean of the
former, meaning the \textit{gap} (difference) between them is
non-negative, regardless of whether we calculate the second-order mean
at the node level and then at the network level, or directly at the
network level. For non-regular graphs, the statement may be
strengthened to a strict inequality.
In this way, the Friendship Paradox establishes a fundamental property of node degrees, which are a key concept in both pure graph theory \cite{diestel-graph-theory-book} and in analyses of the Web and other large networked systems \cite{newman-sirev}.

The Friendship Paradox comes in two related forms, both of which represent
natural ways of formalizing this gap.
In one, the degrees of a node’s friends feature separately in the
final mean; we essentially aggregate over each edge separately.
We refer to this as a \say{list} version because each node contributes a
list of numbers, its friends’ degrees, to the final mean \cite{cantwell2021friendship}.
In the second version, for each node,
we first calculate the average degree of its
friends (the second-order degree for this node),
and then average over those for a final mean
\cite{kramer2016multistep}.
We refer to this as a \say{singular} version because each node contributes
one number, the mean of its friends’ degrees, to the final mean.

It's intuitive to think of degree as a node attribute, and the next
thought is whether \textit{any} numbers assigned as attributes would
produce results similar to when degrees act as attributes. Eom and Jo
looked at just that in their 2014 paper, generalizing the Friendship
Paradox (FP) to the Generalized Friendship Paradox
(GFP)~\cite{eom2014generalized}. They claimed that as long as the
degree sequence and the attribute sample of a non-regular simple
connected finite graph are positively correlated, the gap between the second-order mean and the first
order mean is positive, i.e. the list version of GFP holds. Note that the graph needs to be non-regular for
the degree-attribute correlation to be defined. If either degrees or
attributes of a graph's nodes are all equal, it is undefined. For both regular graphs and graphs where all attributes are equal, the gap would be 0.

The result of Eom and Jo makes concrete an important finding about
network structure, which is that many correlates of degree
seem to satisfy the principle that your friends have \say{more of it
than you do.}
This is relevant in many contexts on the World Wide Web, where different types of user activity are often positively correlated with a user's degree in the underlying social network \cite{de2012not,hodas2013friendship,romero2011influence,zhu2013predicting}.
For example, in an influential early analysis of the Facebook network
as a paradigmatic example of a large web-based social media system,
Ugander et al. showed that attributes such as the time spent on the
site exhibited an analogue of the friendship paradox, and observed
that the correlation between time on site and number of friends
made such a finding natural \cite{ugander2011anatomy}.
The correlation between degree and activity also forms part of the motivation for using individuals' social network degree in immunization strategies \cite{christakis2010social,cohen2003efficient}.
And {\em social comparison theory} \cite{akerlof1997social,burt2010shadow,festinger1954theory} suggests that the greater average degree of one's friends may be a source of comparison \cite{zuckerman2001makes}.

Eom and Jo's claim concerned specifically the list (\say{edge-based,}
\say{network-level}) version of the paradox: as noted above, this is
the version where the attributes of a node's friends feature
separately in the final mean, and not as part of a fraction
representing a node's single-number second-order quantity. Indeed,
suppose that in a simple connected finite graph $G$, node $i$'s degree is $d_i$, its attribute is $a_i$,
the degree sequence has mean $\overline{d} > 0$ and standard deviation (s.d.) $\sigma_d >
0$ indicating non-regularity in degrees, the attribute sample has s.d. $\sigma_a > 0$ indicating non-regularity in attributes, and the
degree-attribute correlation is $r_{d,a}$. Then the list version gap can
be expressed as
$g_{list} = {r_{d,a} \sigma_d \sigma_a} / {\overline{d}.}$

Since $\sigma_d, \sigma_a, \overline{d}$ are all positive, any attribute sample $a$ with positive $r_{d,a}$ satisfies the list version of GFP (LGFP); an attribute sample with zero correlation makes the gap 0, and negative $r_{d,a}$ fails LGFP. In contrast to the regular Friendship Paradox, zero LGFP gap does not make a statement about a graph's regularity, since our domain here is non-regular graphs. 

Given the generality of this result, it is important to understand whether it
also holds for the singular version of the Friendship Paradox,
since that would be the most direct way to formalize the idea that for
any attribute positively correlated with degree, including online activity metrics, \say{your} friends
have more of it than \say{you} do (what's important here is the emphasis on individuals' singular-number second-order values, implying an averaging over nodes).
Our paper starts from the surprising fact that the principle that
attributes positively correlated with degree lead to the gap being positive is false, as a general
statement, for the singular version of the Generalized Friendship Paradox.
Given how naturally the list version of the Friendship Paradox 
generalizes to attributes that are positively correlated with
degree, it's striking to see GFP fail for the singular version.
It's all the more notable since the two versions of the original Friendship Paradox tend to behave very similarly in a mathematical
sense, and to exhibit roughly the same properties, whereas the
failure of this generalization for one version but not the other
drives an unexpected technical wedge between them.

This paper explores what conditions are required for the singular version of GFP to fail and to hold. We will refer to \say{the singular version of GFP} as SGFP. Below, we discuss that for degree-attribute correlations as close to 1 as we want we can find a graph topology and an attribute assignment for which SGFP fails (Section~\ref{sec:path}) but also that each \say{path to correlation 1} goes to infinity in the number of nodes (Section~\ref{sec:path_infinite}). We mention that correlation 1 can't fail SGFP for any topology because the calculations would reduce to those of the Singular Friendship Paradox (full proof in the Supplementary Information). We also show that we can split all graph topologies into those for which we \textit{can} find an attribute sample with positive $r_{d,a}$ (at some value) and for which SGFP fails---so-called anti-SGFP topologies; and those for which we can't find such a sample, meaning that any attribute sample with $r_{d,a}>0$ leaves SGFP standing---pro-SGFP topologies (Section~\ref{sec:farkas}). We also explore how often random networks are pro-SGFP (as the number of nodes grows they are less and less common) and how high SGFP-failing correlations may be for random anti-SGFP networks (Section~\ref{sec:pro_counting}). Finally, we turn to real-world data and see that real attributes may fail SGFP (Section~\ref{sec:real}). We also find how non-trivially large SGFP-failing correlations may be for real social networks and explore what happens if we remove social structure by rewiring the graphs (Section~\ref{sec:real_optimize}).

\section{Definitions}
\label{sec:def}
Our domain is simple connected finite graphs. We are looking at connected graphs, but similar to the Friendship Paradox, our results are easily generalized to disconnected graphs including those with isolates. For a graph with disjoint connected components of size $\geq 2$, the gap can be calculated node-by-node without regard for disconnectedness. (Note that while graph regularity plays a role in our discussion, it would generalize to graph regularity \textit{within each component} in the case of multiple components). If there are isolated nodes whose second-order attributes are undefined, we can simply disregard them. When dealing with real data, we follow these principles to calculate the gaps. For the theoretical part of this paper, we assume connectedness so we can rule out isolates and iterate over all node indices when computing various values.

For a node $i$ in such a graph on $n$ nodes, let $d_i$ be its degree and $a_i$ be its attribute for some attribute assignment $a$ (we use terms \say{attribute sample} and \say{attribute assignment} interchangeably). Let $s_i$ be $i$'s singular second-order attribute, i.e. the single-number mean of its friends' attributes. Let the degree sequence have mean $\overline{d}$, the attribute sample have mean $\overline{a}$, and the set of second-order attributes have mean $\overline{s}$. Let the gap $g$ be equal to the difference between the mean second-order attribute and the mean first-order attribute, so
$$g = \frac{1}{n}\sum_{i=1}^{n} s_i - \frac{1}{n}\sum_{i=1}^{n} a_i $$
This can also be written more compactly as
$g = \overline{s} - \overline{a} $,
or in more detail as
\begin{equation}
\label{eqn:gap}
g = \frac{1}{n}\sum_{i=1}^{n} \left(\frac{1}{d_i} \sum_{j \in N(i)} a_j \right) - \frac{1}{n}\sum_{i=1}^{n} a_i 
\end{equation}
where $N(i)$ is the set of neighbors, or friends, of $i$.
It will also be useful to express the gap differently. In the first term of equation~\ref{eqn:gap}, index $i$ refers to nodes (\textit{seeds}) and index $j$ to $i$'s friends, and we look at the seeds' calculations of their second-order attributes. Here, we do the opposite and see how each node $j$'s attribute features in its friends' second-order values. $j$'s coefficient in the second-order mean is equal to $\frac{1}{n}$ times the sum of $j$'s friends' reciprocal degrees: 
$$ g = \frac{1}{n}\sum_{j=1}^{n} a_j \left( \sum_{k \in N(j)} \frac{1}{d_k} \right) - \frac{1}{n}\sum_{j=1}^{n} a_j$$
where $N(j)$ is the set of $j$'s friends. We call $\sum_{k \in N(j)} \frac{1}{d_k}$, the sum of $j$'s friends' reciprocal degrees, $\delta_j$, and could also write:
\begin{equation}
\label{eqn:gap2}
g = \frac{1}{n}\sum_{j=1}^{n} \delta_j a_j - \frac{1}{n}\sum_{j=1}^{n} a_j
\end{equation}

We say that SGFP fails if $g<0$ and holds otherwise.

If all degrees or all attributes in a graph are the same, the degree-attribute correlation is undefined. In each of these cases, the gap is 0, so SGFP doesn't fail. See Supplementary Information for a proof.

While graphs that exhibit regularity in degrees or attributes produce a gap of 0, these are not the only cases when that happens (See the end of Section~\ref{sec:farkas} for details). For SGFP, it's important not to think of a gap of 0 as implying a graph's regularity. As a simple example, consider a path graph $x-y-z$ with $a_x$, $a_y$ and $a_z$ as attributes. The gap is equal to $\frac{1}{3} \left(\frac{a_x+a_z}{2} + a_y + a_y \right) - \frac{1}{3}\left(a_x+a_y+a_z \right)=-\frac{a_x}{6}+\frac{a_y}{3}-\frac{a_z}{6}$ which is equal to 0 if $a_x+a_z=2a_y$. So a path graph $x-y-z$ with 1, 2, and 3 as attributes is neither regular nor has the same attributes but produces a gap of 0.

Having dealt with the \say{regular} cases, below we will further assume that our simple connected finite graphs are non-regular and that for each attribute sample, not all its values are the same, so the correlation between the degree sequence and the attribute sample is defined. It is a measure of linear dependence between them and it is given by
\begin{equation}
 r_{d,a} = \ddfrac{\sum_{i=1}^{n} (a_i - \overline{a}) (d_i - \overline{d})}{\sqrt{\sum_{i=1}^{n} (a_i - \overline{a})^2 \sum_{i=1}^{n} (d_i - \overline{d})^2 }}
 \label{eqn:corr}
\end{equation}

\section{A path to correlation 1}
\label{sec:path}
We claim that for any $\epsilon > 0$, there is a graph topology and an attribute assignment such that the degree-attribute correlation $r_{d,a} > 1-\epsilon$ and SGFP fails. In this section, we construct one example of a set of graphs such that for any $\epsilon > 0$, this specific set contains a graph that fits the requirements of $r_{d,a} > 1-\epsilon$ and SGFP failing.

\begin{figure}[h]
\centering
\includegraphics[width=0.4\textwidth]{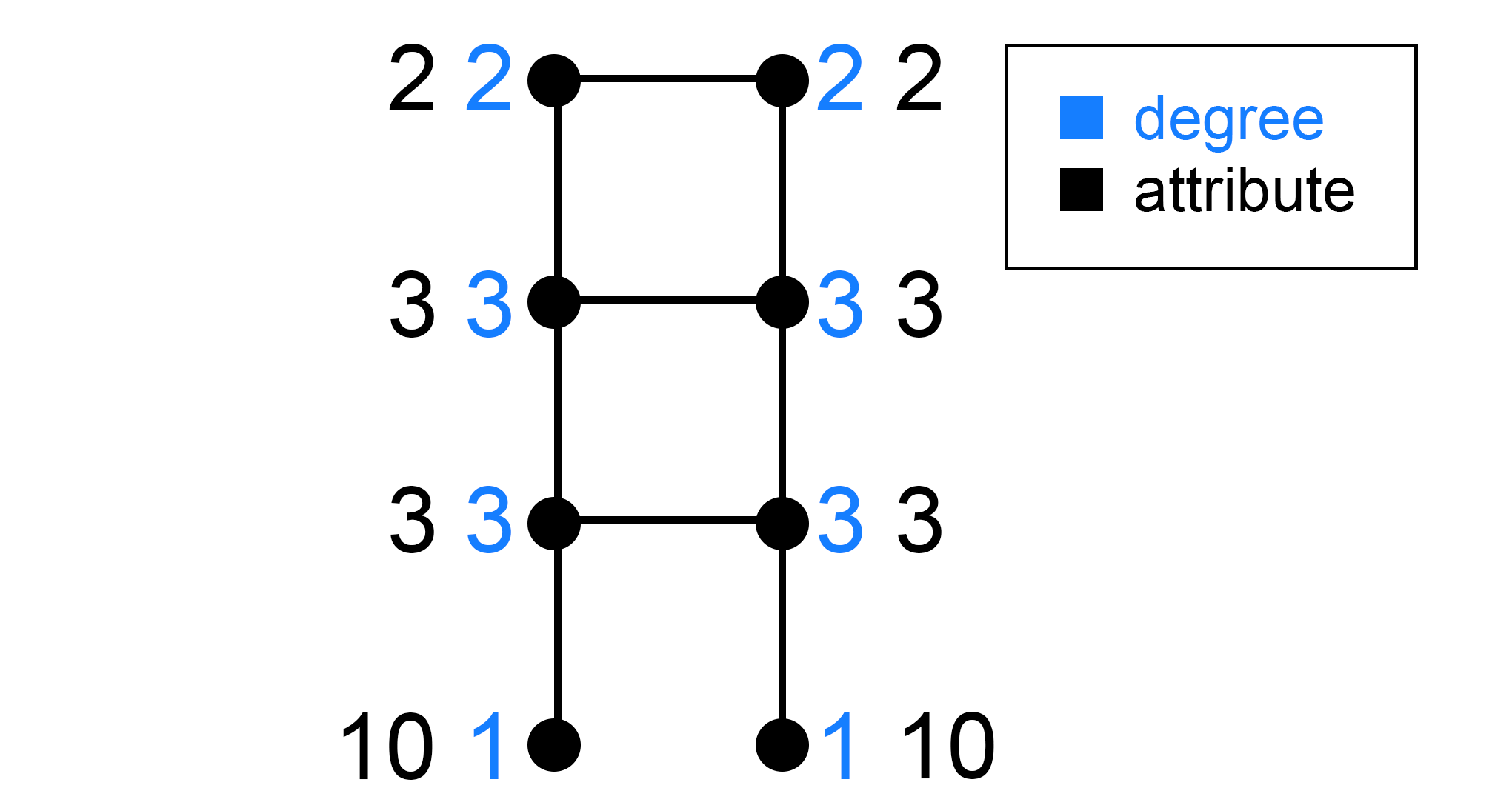}
\caption{Starting with an SGPF-failing graph such as this, we could grow it by adding nodes so that each new graph still fails SGFP and the degree-attribute correlation approaches 1.
\label{fig:example}
}
\end{figure}

Take a graph with $n$ nodes and some attribute assignment for which SGFP fails, meaning the gap $g$ is negative. Take a look at an example in Figure~\ref{fig:example}.

For the graph in Figure~\ref{fig:example}, the gap is equal to:
$$g = \overline{s}-\overline{a} = \frac{1}{8} ( \frac{1}{2}(2+3)+\frac{1}{2}(2+3)+
\frac{1}{3}(2+3+3)+\frac{1}{3}(2+3+3)+$$ 
$$\frac{1}{3}(3+3+10)+\frac{1}{3}(3+3+10)+
\frac{1}{1}(3)+\frac{1}{1}(3)) -$$$$ \frac{1}{8}\left(2+2+3+3+3+3+10+10 \right) = -\frac{9}{8} < 0$$

The degree-attribute correlation for that graph is $-\frac{17}{\sqrt{451}}$ $\approx -0.80$.

\begin{figure}[h]
\centering
\includegraphics[width=0.4\textwidth]{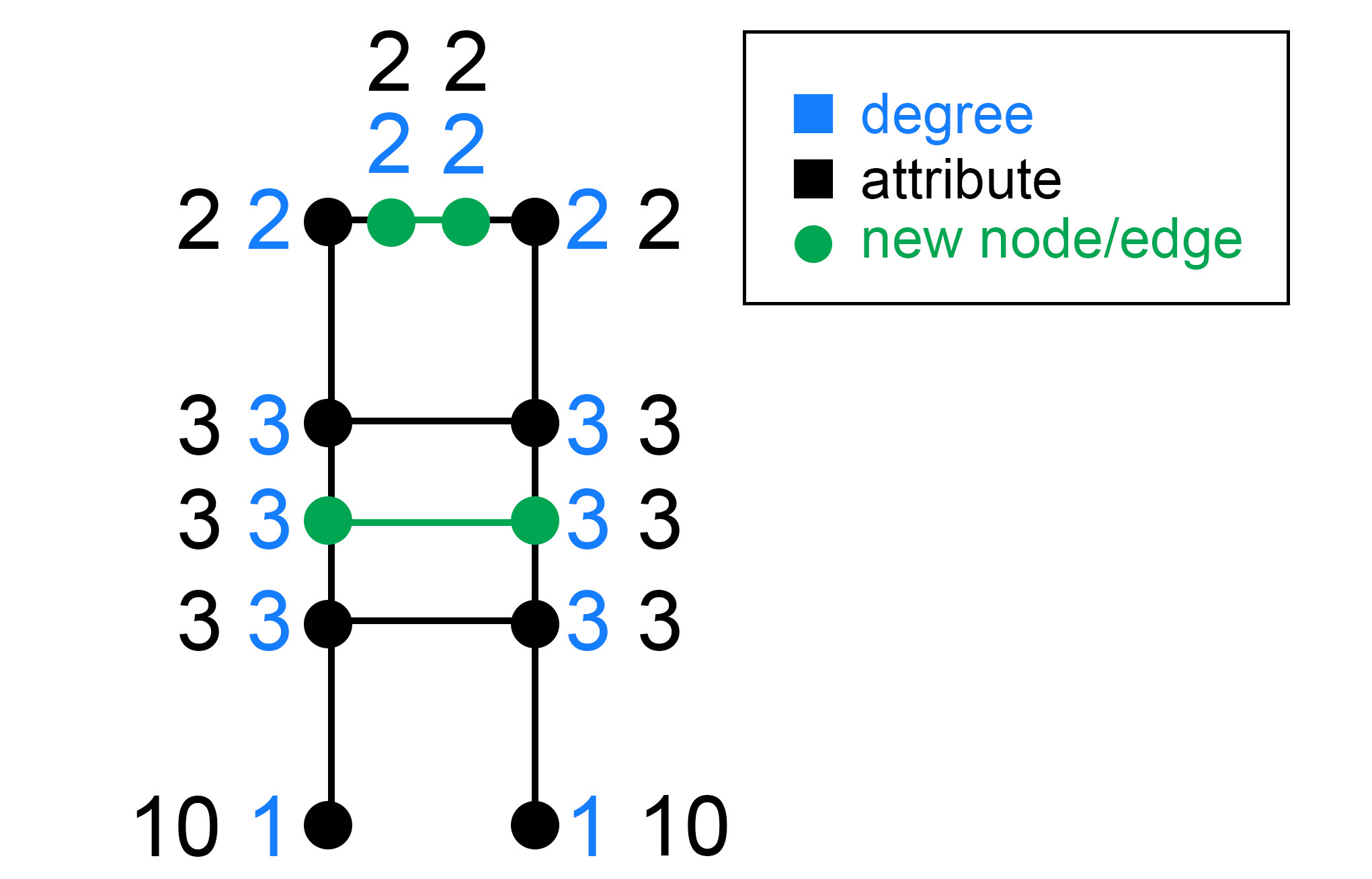}
\caption{Adding nodes in a specific way preserves the Singular Generalized Friendship Paradox gap sign and increases the degree-attribute correlation to the limit of 1. This figure is an example of a first step of this process, with new nodes and edges in green.
\label{fig:add}
}
\end{figure}

Now, suppose we add 4 new nodes to the graph: two nodes with degree 2, attribute 2 and second-degree attribute 2 (\say{triple-2} nodes) and two nodes with degree 3, attribute 3 and second-degree attribute 3 (\say{triple-3} nodes) while keeping the degrees, attributes and \textit{second-order} attributes of the original nodes the same. For the graph in Figure~\ref{fig:example}, we can perform the first step as shown in Figure~\ref{fig:add}.

If a triple-2 node $i$ comes between two nodes with attribute 2, its second-order attribute $s_i$ will indeed be 2, and its friends' second-order attributes will not be affected, $i$ effectively taking the place of the former attribute-2 friend for each of \textit{its} friends. Similarly, adding triple-3 nodes in a certain way ensures their second-order attribute is 3 and preserves the second-order attributes of its friends.

Here, it will help us to rewrite the gap yet another way:
$$g = \frac{1}{n}\sum_{i=1}^{n} \left(s_i - a_i\right) $$
Since each new node $p$ has $s_p = a_p$, the new node adds nothing to the sum. It does, however, increase the number of nodes $n$, so the gap remains negative but decreases in absolute value. (While we don't discuss this in detail, it is possible to counter the gap's convergence to 0 from below with a slightly more elaborate way of adding nodes). 

Now that we know how to add 4 nodes to this graph and not change the sign of its gap, let's call \say{adding 4 nodes} a step of induction, since we can perform the same procedure to the graph in Figure~\ref{fig:add} and so on. By induction, after each step, the gap is still negative. Let's number the initial Figure~\ref{fig:example} nodes 1 through 8. After $k$ steps, with $2k$ triple-2 nodes and $2k$ triple-3 nodes added, we have $n=8+4k$ nodes, and the degree-attribute correlation is given by
\small
$$r^{(k)} = \ddfrac{\sum_{i=1}^{8} (a_i - \overline{a}^{(k)}) (d_i - \overline{d}^{(k)}) + 2k (2-\overline{a}^{(k)})(2-\overline{d}^{(k)}) + 2k (3-\overline{a}^{(k)})(3-\overline{d}^{(k)})}{\sqrt{\sum_{i=1}^{8} (a_i - \overline{a}^{(k)})^2 + 2k (2-\overline{a}^{(k)})^2 + 2k (3-\overline{a}^{(k)})^2}} \times$$$$ \ddfrac{1}{\sqrt{\sum_{i=1}^{8} (d_i - \overline{d}^{(k)})^2 + 2k (2-\overline{d}^{(k)})^2 + 2k (3-\overline{d}^{(k)})^2}}$$
\normalsize
with $\overline{a}^{(k)}$ and $\overline{d}^{(k)}$ being the means of the graph's attribute sample and the degree sequence respectively after $k$ steps.
We're interested in the limit of $r^{(k)}$ as $k$ goes to infinity. Since $\lim_{k\to\infty} \overline{d}^{(k)}=\lim_{k\to\infty} \overline{a}^{(k)}=2.5$, $\lim_{k\to\infty} r^{(k)}$ equals

$$\lim_{k\to\infty} \ddfrac{\sum_{i=1}^{8} (a_i - 2.5) (d_i - 2.5) + 4k \cdot 0.5^2}{\sqrt{\left(\sum_{i=1}^{8} (a_i - 2.5)^2 + 4k \cdot 0.5^2\right) \left( \sum_{i=1}^{8} (d_i - 2.5)^2 + 4k \cdot 0.5^2 \right)}} $$

The sums are finite numbers which we can write as constants:
$$\lim_{k\to\infty} r^{(k)} = \lim_{k\to\infty} \ddfrac{C_1 + k}{\sqrt{\left(C_2 + k \right) \left( C_3 + k \right)}} =$$

$$ \lim_{k\to\infty} \ddfrac{C_1 + k}{\sqrt{C_2 C_3 + C_2 k + C_3 k + k^2 }} = 1 $$

So, adding 2 triple-2 and 2 triple-3 nodes at a time preserves the negative gap and grows the correlation to 1 in the limit. For different constructions, $r_{d,a}$ may surpass $1-\epsilon$ in fewer or more steps depending on the degree-attribute correlation in the original graph, purposely taken to be very low in our example. It is also possible to counter the gap's convergence to 0, instead keeping the gap constant at each step. But since our goal is to provide \textit{an example} path to 1 where the gap is \textit{negative} at each step, not large in absolute value, we won't go into detail about those techniques here. For now, we have shown that for any $\epsilon > 0$, there is a graph topology and an attribute assignment such that the degree-attribute correlation $r_{d,a} > 1-\epsilon$ and the gap is negative.

\begin{figure}[h]
\centering
\includegraphics[width=0.4\textwidth]{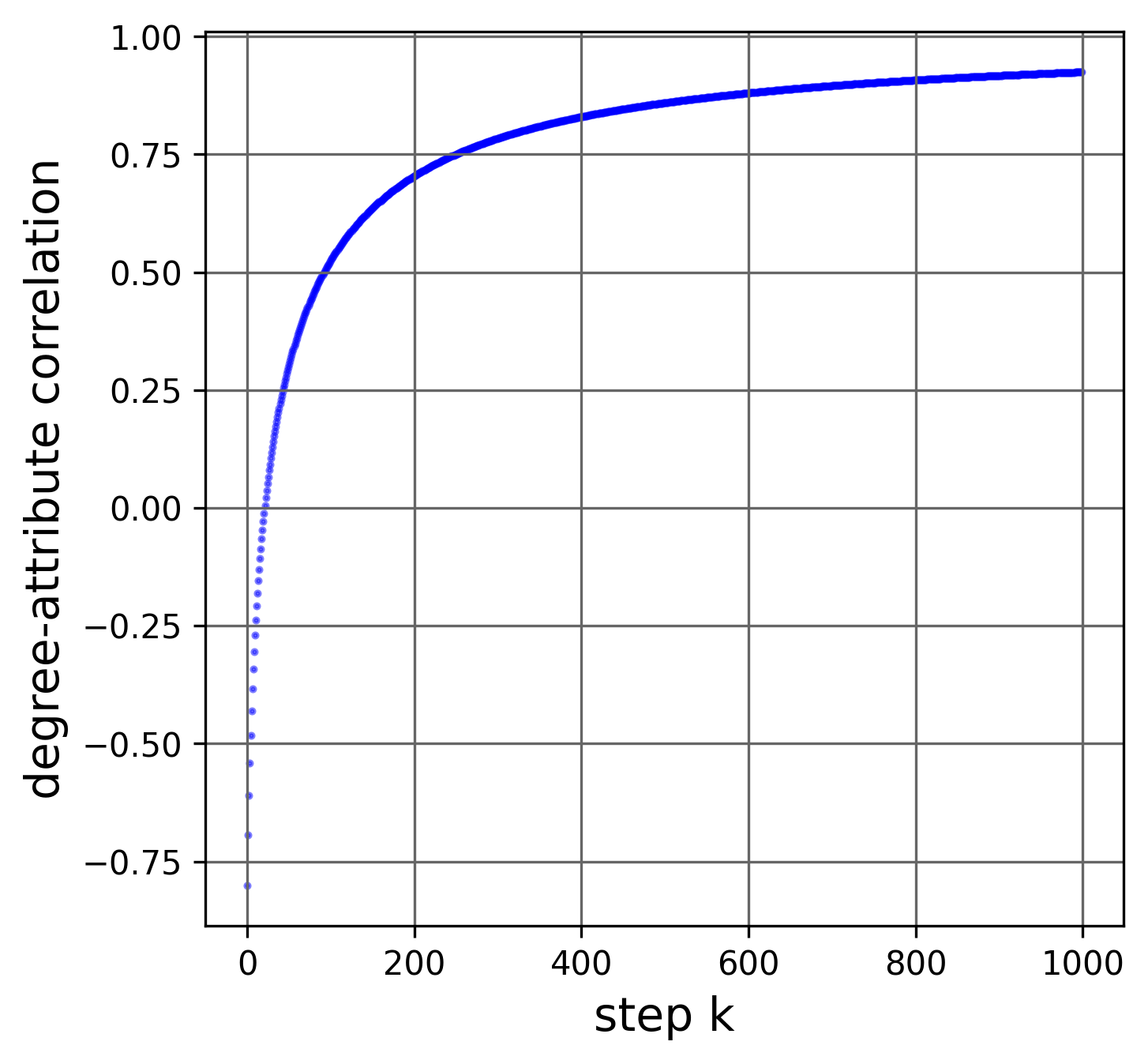}
\caption{Degree-attribute correlation for the 8-node example in Figure~\ref{fig:example} as we add pairs of triple-2 and triple-3 nodes.
\label{fig:corr_plot}
}
\end{figure}

The correlation for the graph in Figure~\ref{fig:example} is shown in Figure~\ref{fig:corr_plot}.

The path to 1 that we've constructed has 1 as the limit, not a destination that is reached, because an attribute sample with $r_{d,a}=1$ can't fail SGFP. In the case of $r_{d,a}=1$, the calculation of the SGFP gap simplifies to that of the GFP gap, and we know that GFP holds for all non-regular connected graphs, which is our domain here. Please see the Supplementary Information for a detailed proof.

\section{Each path to 1 goes to infinity in the number of nodes}
\label{sec:path_infinite}

Our earlier construction produced a sequence of graphs and attribute samples whose correlation with the degree sequence converged to 1, and where the gap was negative at each step.
In this sequence, the size of the graphs became arbitrarily large, and it's natural to ask if this is necessary; is it possible that for some fixed finite graph topology $G$, there is a sequence of attribute samples with negative gaps and the degree-attribute correlation converging to 1?

In this section we show that this is not possible, because we are able
to show the following result: for every finite graph $G$, there is a
constant $\varrho_G < 1$ such that any attribute sample in $G$ whose
correlation with the degree sequence exceeds $\varrho_G$ must produce
a non-negative gap. Thus, every graph has a threshold strictly below
1 such that correlations above this threshold cannot fail SGFP.

To show this, we'll consider a graph $G$ with degree sequence $d$. An \textit{attribute sample} $a$ is
the set of attributes of nodes in $G$, indexed in the same way as the
degrees, so that a degree-attribute correlation $r_{d,a}$ is well-defined. A \textit{sequence of attribute samples}
$a^{(k)}$ is a collection of such samples, and for our purposes
$r_{d,a^{(k)}}$ converges to 1. We denote individual nodes' attributes
$a_i^{(k)}$. We'll now establish that the gaps for these attribute
samples must eventually become non-negative as we move through the
sequence.

We start with the following fact:
\begin{quote}
{($\ast$) \em For every $\epsilon > 0$ there exists an $\alpha > 0$ such that
the following holds.
If $x = (x_1, ..., x_n)$ and $y = (y_1, ..., y_n)$ are vectors
for which $\sum_{i=1}^n x_i^2 = \sum_{i=1}^n y_i^2 = 1$ and
$\sum_{i=1}^n x_i y_i > 1 - \alpha$, then 
$|x_i - y_i| < \epsilon$ for all $i$.}
\end{quote}
We prove this fact with $\alpha = \epsilon^2/2$.
Using this as the value of $\alpha$ in terms of $\epsilon$, we first
observe that 
\begin{eqnarray*}
\sum_{i=1}^n (x_i - y_i)^2 & = &
\sum_{i=1}^n (x_i^2 - 2 x_i y_i + y_i^2) \\
& = &
\sum_{i=1}^n x_i^2 + \sum_{i=1}^n y_i^2 - 2 \sum_{i=1}^n x_i y_i \\
& < & 2 - 2(1 - \alpha) = 2 \alpha = \epsilon^2.
\end{eqnarray*}
Thus for each $i$ from $1$ to $n$ we have
$$(x_i - y_i)^2 \leq \sum_{i=1}^n (x_i - y_i)^2 < \epsilon^2,$$
and taking square roots we have
$|x_i - y_i| < \epsilon$, which proves ($\ast$).

Now, fix a non-regular graph $H$ on $n$ nodes.
For any $n$-dimensional vector $a = (a_1, ..., a_n)$, we can view
it as an assignment of attribute values to the nodes of $H$.
Let $g_H(x)$ be the gap for this vector of attributes under the
gap definition in the paper.
For the vector of degrees $d = (d_1, ..., d_n)$, the gap
$g_H(d)$ is a constant $g_H > 0$, by the singular version of the
Friendship Paradox.

$g_H(a)$ is unaffected if we add the same
constant to each coordinate of $a$ (see Supplementary Information for the proof).

This means we can subtract the mean of $a$ from each $a_i$; and if the sum of the coordinates of $a$ is now $0$, then, following equation~\ref{eqn:gap2}:
$$g_H(a) = \frac1n \sum_{i=1}^n \delta_i a_i$$
Applying this to the vector of degrees, let
$\overline{d}$ be the average degree in $H$;
let $\omega = \sqrt{\sum_{i=1}^n (d_i - \overline{d})^2}$;
and let $\tau_i = (d_i - \overline{d})/\omega$.
Then the vector $\tau = (\tau_1, ..., \tau_n)$ is a unit vector
whose coordinates sum to $0$; that is, $\sum_{i=1}^n \tau_i^2 = 1$
and $\sum_i \tau_i = 0$.
And since $\tau = d/\omega - \overline{d}/\omega$ for
constants $\omega$ and $\overline{d}$,
we have 
$$g_H(\tau) = \frac1n \sum_{i=1}^n \delta_i \tau_i 
= g_H(d)/\omega = g_H/\omega > 0$$

Now, suppose we have a sequence of attribute vectors 
$\{a^{(k)} : k = 1, 2, 3, ... \}$ such that 
$g_H(a^{(k)}) < 0$ for all $k$, and the correlation of
$a^{(k)}$ with $d$ converges to $1$ as $k$ goes to infinity.
Since the correlation of two vectors
is unaffected if we subtract the same constant from each coordinate
of one of the vectors, and multiply the vector by a positive coefficient,
we can assume without loss of generality that
each $a^{(k)}$ is a unit vector whose coordinates sum to $0$.
Applying the same argument to transform $d$ to $\tau$, it follows
that the correlation of the vectors $a^{(k)}$ with $\tau$ 
converges to $1$ as $k$ goes to infinity.

For two unit vectors $x$ and $y$ whose coordinates each sum to $0$, 
their correlation is simply their inner product, 
since in this case the numerator of the expression for correlation
is $\sum_{i=1}^n x_i y_i$ and the denominator is the square root of
$\left(\sum_{i=1}^n x_i^2\right)\left(\sum_{i=1}^n y_i^2\right) = 1$.
Therefore, the correlation of $a^{(k)}$ and $\tau$ is their
inner product $\sum_{i=1}^n a_i^{(k)} \tau_i$.

Let $\delta^*$ be the maximum value of $\delta_i$,
the coefficients in the expression for the gap.
We now apply our initial fact ($\ast$) with $\epsilon = g_H/(\omega \delta^*)$,
and we obtain the corresponding $\alpha = \epsilon^2$.
Since the correlation of $a^{(k)}$ and $\tau$ converges to $1$
as $k$ goes to infinity, there is a value of $k$ for which 
this correlation is greater than $1 - \alpha$; that is,
$\sum_{i=1}^n a_i^{(k)} \tau_i > 1 - \alpha$.

In this section we show that this is not possible, because we are able
to show the following result: for every finite graph $G$, there is a
constant $\varrho_G < 1$ such that any attribute sample in $G$ whose
correlation with the degree sequence exceeds $\varrho_G$ must produce
a non-negative gap. Thus, every graph has a threshold strictly below
1 such that correlations above this threshold cannot fail SGFP.

We now show that that any attribute sample in $G$ whose
correlation with the degree sequence exceeds $1 - \alpha$ must produce
a non-negative gap.
By fact ($\ast$) for such a $k$ where
the correlation is greater than $1 - \alpha$, we have 
$|a_i^{(k)} - \tau_i| < \epsilon$ for all $i$.

For such a $k$, we have
$$g_H(\tau) - g_H(a^{(k)}) 
= \frac1n \sum_{i=1}^n \delta_i (\tau_i - a_i^{(k)})
< \frac1n \sum_{i=1}^n \delta_i \epsilon
=$$$$\frac1n \sum_{i=1}^n \delta_i {g_H} / (\omega \delta^*)
\leq \frac1n \cdot n g_H / \omega
= g_H / \omega,$$
and thus $g_H(\tau) - g_H(a^{(k)}) < g_H / \omega.$
But this contradicts the fact that 
$g_H(\tau) = g_H/\omega$ and $g_H(a^{(k)}) < 0$.
This contradiction establishes that there can't be such a
sequence $a^{(k)}$ that produces a negative gap and
whose correlation with $\tau$ (and hence with $d$) converges to $1$ for a finite graph $H$.

Given this result, we can consider 
the set $S_G$ of all $\beta$ with
the property that any attribute sample in $G$ whose
correlation with the degree sequence exceeds $\beta$ must produce
a non-negative gap. 
This set has an infimum, which we can denote by $\varrho_G$. It is a property of topology $G$ such that degree-attribute correlations at or below $\varrho_G$ might fail SGFP, but degree-attribute correlations above it can't.

\section{Differentiating between pro- and anti-SGFP graphs}
\label{sec:farkas}

Given the discussion so far, it is natural to ask which graphs $G$ have the property that \textit{there exists} an attribute sample $a$ on $G$ that fails SGFP while $r_{d,a}>0$.
We will call such a graph an {\em anti-SGFP} topology (since it is capable of refuting SGFP); and if a graph does not have this property, we will call it {\em pro-SGFP}. 
For pro-SGFP topologies, $r_{d,a}>0$ would imply a positive gap, $r_{d,a}=0$ a gap of 0, and $r_{d,a}<0$ a negative gap, which is the case for all non-regular connected topologies with the list version of GFP.

Is there a tractable characterization of the anti-SGFP and pro-SGFP graphs?
We establish here that there is, through a clean characterization showing whether a graph is anti-SGFP or pro-SGFP. 

First, consider a graph and an attribute assignment. Recall from Supplementary Information that adding a constant $c$ to each attribute doesn't change the gap. It also doesn't change the correlation, since in the correlation formula we subtract $\overline{a}$ from each $a_i$, and if each $a_i$ grows by $c$, $\overline{a}$ grows exactly by $c$. Thus, for any SGFP-failing attribute assignment, we can change it so its mean $\overline{a}$ is 0 (by subtracting the original mean from each attribute value) and retain the SGFP-failing property. Meaning, if we can find an attribute assignment $a'$ with $r_{d,a'}>0$ that fails SGFP, we can find an attribute assignment $a$ with all the same characteristics and an additional constraint $\overline{a} = 0$.
Then, our gap formula becomes:
$$g(\overline{a}=0) = \frac{1}{n}\sum_{i=1}^{n} \left( \frac{1}{d_i} \sum_{j \in N(i)} a_j \right)$$
or, given our discussion of $\delta_i$ above:
$$ g = \frac{1}{n}\sum_{i=1}^{n} \delta_i a_i $$
where $\delta_i = \sum_{j \in N(i)} \frac{1}{d_j} $.

The correlation becomes:

$$r_{d,a} (\overline{a}=0) = \ddfrac{\sum_i a_i (d_i - \overline{d})}{\sqrt{\sum_i a_i^2 \sum_i (d_i - \overline{d})^2 }} $$

The correlation is positive if and only if its numerator is positive:
$$ \sum_i a_i (d_i - \overline{d}) > 0$$
From here, we get
$$ \sum_i a_i d_i - \sum_i a_i \overline{d} = 
\sum_i a_i d_i - \overline{d} \sum_i a_i =
\sum_i a_i d_i - \overline{d} 0 =
\sum_i a_i d_i > 0
$$
So, if we have an attribute assignment $a$ with mean 0 and $r_{d,a}>0$ and that assignment fails SGFP, we have 3 constraints:
\begin{equation}
 \sum_i a_i = 0
 \label{eqn:constraint_a}
\end{equation}
\begin{equation}
 \sum_i d_i a_i > 0
 \label{eqn:constraint_da}
\end{equation}
\begin{equation}
 \sum_i \delta_i a_i < 0
 \label{eqn:constraint_g}
\end{equation}

Then, given a graph $G$, we want to see if we can find an attribute assignment that satisfies the 3 constraints above. If we can, it's an anti-SGFP graph, and if we can't, it's a pro-SGFP graph.

Before we go further, we want to show that if we can satisfy $\sum_i a_i = 0$, $\sum_i \delta_i a_i < 0$ (negative gap) and $\sum_i d_i a_i = 0$ (\textbf{zero} correlation) for a graph $G$, we will also be able to satisfy $\sum_i d_i a_i > 0$ (positive correlation). Let's say that for an attribute sample $a$ with $r_{d,a}=0$, $g_0 = \frac{1}{n} \sum_i \delta_i a_i = x < 0$ meaning SGFP fails. Let node $i$ be the highest-degree node, or one of them, and node $j$ be the lowest-degree node, or one of them. Since the graph is non-regular, we know $d_i>d_j$. We will raise $a_i$ by $\epsilon > 0$ and decrease $a_j$ by $\epsilon$. $\overline{a}$ will remain at 0, the change in $\sum_i d_i a_i$ will be $\epsilon(d_i- d_j) >0 $ so we satisfy inequality~\ref{eqn:constraint_da} and get $r_{d,a}>0$. The change in the gap will be $\frac{1}{n} \epsilon(\delta_i - \delta_j)$. In order to keep the gap negative and inequality~\ref{eqn:constraint_g} satisfied, we need $\frac{1}{n} \epsilon(\delta_i - \delta_j) < |x|$. If $\delta_i \leq \delta_j$, we are good with any $\epsilon > 0$. If $\delta_i > \delta_j$ we choose an $\epsilon$ such that
$$\epsilon < \frac{n|x|}{(\delta_i - \delta_j)}$$
So, if there is an SGFP-failing attribute assignment with $r_{d,a}=0$, there is also one with $r_{d,a}>0$. This means that we can replace constraint
$$ \sum_i d_i a_i > 0 $$
with 
\begin{equation}
 \sum_i d_i a_i \geq 0
 \label{eqn:constraint_da_new}
\end{equation}
We can also replace 
$$\sum_i a_i = 0$$
with the equivalent pair of inequalities 
\begin{equation}
 \sum_i a_i \geq 0
 \label{eqn:constraint_ap}
\end{equation}
\begin{equation}
 \sum_i -a_i \geq 0
 \label{eqn:constraint_am}
\end{equation}

The final system has 4 inequalities
\begin{equation}
\label{eqn:system}
\begin{cases}
\begin{aligned}
\sum_i a_i \geq 0 \\
\sum_i -a_i \geq 0 \\
\sum_i d_i a_i \geq 0 \\
\sum_i \delta_i a_i < 0

\end{aligned}
\end{cases}
\end{equation}

We will use Farkas's Lemma to deal with them \cite{Farkas}.
\begin{quote}
{\em Farkas's Lemma: Let $A$ be a matrix and $b$ a vector.
There exists a vector $x \geq 0$ satisfying $Ax = b$ if and only if
there does not exist a vector $y$ satisfying $yb < 0$ and $yA \geq 0$.}
\end{quote}

Our system fits naturally into the format
of the \say{$y$} system in Farkas's Lemma:
we let $y$ be the vector $(a_1, a_2, \ldots, a_n)$,
$b$ be the vector $(\delta_1, \delta_2, \ldots, \delta_n)$,
and $A$ be a matrix with $n$ rows and $3$ columms, where
the first column is $(d_1, d_2, \ldots, d_n)$,
the second column has all 1's, and the third column has all $-1$'s,
Then our system of inequalities is indeed 
$yb < 0$ and $yA \geq 0$.

By Farkas's Lemma, this system has no solution 
(in other words, the graph $G$ is a pro-SGFP graph)
if and only if the system $Ax = b$ has a solution with $x \geq 0$.
What would this mean?
Since $A$ has only three columns, $x$ is a 3-dimensional vector:
$x = (x_1, x_2, x_3)$.
The $i^{\rm th}$ row of $Ax = b$ corresponds to the 
equation $d_i x_1 + x_2 - x_3 = \delta_i$.

Now, it's unnecessary to have to write $x_2 - x_3$, where
$x_2$ and $x_3$ are both non-negative in each of these $n$ equations (one for each $i = 1, 2, \ldots, n$).
Instead, we can define the variable $z = x_2 - x_3$ and
notice that $z$ can be an arbitrary number, not necessarily
non-negative, since it's equal to one arbitrary non-negative
number minus another one, and hence can be anything at all. Let's also rewrite $x_1$ as $x$.

Therefore, 
\begin{quote}
{($\ast\ast$) \em Our system $yb < 0$ and $yA \geq 0$ has no solution
if and only if we can find two numbers, $x \geq 0$ and $z$ 
(unconstrained in sign), so that 
$d_i x + z = \delta_i$ for each $i$.}
\end{quote}

A graph $G$ is a pro-SGFP graph if and only if two such
numbers $x$ and $z$ exist; otherwise, it is an anti-SGFP graph. 

Note that if all $\delta_i$ in a connected graph are equal, it is a regular graph (see Supplementary Information for the proof). Our domain is non-regular connected graphs, so we know both $d_i$ and $\delta_i$ have variation. This means that for all graphs in our domain, the degree-delta correlation $r_{d,\delta}$ is defined.

For pro-SGFP graphs, $x$ in ($\ast\ast$) can't be equal to 0 (since that would mean all $\delta_i$ are the same), and so $x>0$. Complying with ($\ast\ast$) in this case means that $r_{d,\delta}=1$. So, 
\begin{quote}
\em A (non-regular connected) graph is pro-SGFP if its $r_{d,\delta}=1$\\ and anti-SGFP otherwise.
\end{quote}

Note that for a graph to have two nodes $i$ and $j$ such that $d_i = d_j \textrm{ but } \delta_i \neq \delta_j$ would be sufficient (but not necessary) for a graph to be anti-SGFP. That means that attaching a path $p-q-r-s-t$ to \textit{any} graph at node $p$ would make said graph anti-SGFP: nodes $r$ and $s$ would both have degree 2, but $\delta_r = \frac{1}{d_q}+\frac{1}{d_s}=\frac{1}{2}+\frac{1}{2}=1$ and $\delta_s=\frac{1}{d_r}+\frac{1}{d_t}=\frac{1}{2}+\frac{1}{1}=1.5$. Attaching two leaf nodes to different-degree nodes of any graph would make that graph anti-SGFP as well since the two leaf nodes would have different $\delta_i$. 

It's clear that regular graphs (which are not part of our domain, don't have defined $r_{d,\delta}$, and aren't characterized as pro-SGFP) satisfy the property ($\ast\ast$) above, and it's interesting to think of that property as being part of the regularity of regular graphs. Then, we could think of pro-SGFP graphs as \say{semi-regular,} where they satisfy the property ($\ast\ast$) (the first requirement for being a regular graph) but do not actually have all nodes sharing degrees (the second requirement for being a regular graph).

For a pro-SGFP graph $G$, the fact that we can't find an attribute sample $a$ with $r_{d,a}>0$ and a negative gap means that for all possible attribute samples $a$ such that $r_{d,a}>0$, the gap is non-negative. We can strengthen that to each gap being positive. 

Suppose we have a pro-SGFP graph $G$ and an attribute sample $a$ such that $r_{d,a}>0$. As shown above, we can alter $a$ to have mean 0 and keep $r_{d,a}$ and the gap the same. Then we have 4 constraints:
\begin{equation*}
\begin{cases}
\begin{aligned}
\sum_i a_i = 0 \\
\sum_i d_i a_i > 0 \\
\forall i\textrm{, }d_i x + z = \delta_i\\
x>0
\end{aligned}
\end{cases}
\end{equation*}
with the second constraint satisfying the positive correlation requirement and the third constraint indicating the graph being pro-SGFP. The sign of $\sum_i \delta_i a_i$ would indicate the sign of the gap. Expressing $\sum_i \delta_i a_i$ in terms of $d_i$ given the third constraint, we get
$$\sum_i \delta_i a_i = \sum_i (d_i x + z) a_i$$ 
and finally
$$\sum_i \delta_i a_i = x \sum_i d_i a_i + z \sum_i a_i$$
The first term is positive by the multiplication of constraints 2 and 4. The second term is 0 by the first constraint. This means that $\sum_i \delta_i a_i$ is positive and for pro-SGFP topologies, $r_{d,a}>0$ implies specifically a positive gap.
Changing the sign of the second constraint to indicate zero and negative $r_{d,a}$, we can also see that for $r_{d,a}=0$, the gap is 0, and for $r_{d,a}<0$, it is negative.
Overall, this means that 
\begin{quote}
\em For pro-SGFP topologies, the sign of $r_{d,a}$ determines the sign of the SGFP gap.
\end{quote}

In their discussion of the list version of GFP, Eom and Jo noted that for all non-regular connected topologies, the sign of $r_{d,a}$ determined the sign of the LGFP gap \cite{eom2014generalized}. This means that in our vocabulary, all non-regular connected topologies are pro-LGFP. This is starkly different from SGFP, where there is a further strict constaint of $r_{d,\delta}=1$.

Simple examples of pro-SGFP graphs are \say{star} graphs, where one node is connected to many leaf nodes. Consider also that for each $n$, there is one complete graph $K_n$, and ${n \choose 2}$ labeled networks $K_n - e$, complete graphs missing an edge. In fact, each $K_n - e$ (\say{knee}) graph is pro-SGFP, as it only has two kinds of degrees ($n-1$ and $n-2$), and nodes that share degrees also share $\delta_i$ due to symmetry, with $r_{d,\delta}=1$.

Another thing to note is that a pro-SGFP topology with $r_{d,a}=0$ is an example of a non-regular graph with differing attributes producing a gap of 0, the value of 0 not indicating regularity (we mentioned that there are such cases in Section~\ref{sec:intro}, and in fact, the 3-node path we described there is pro-SGFP). 

\begin{figure}[t]
\centering
\includegraphics[width=0.4\textwidth]{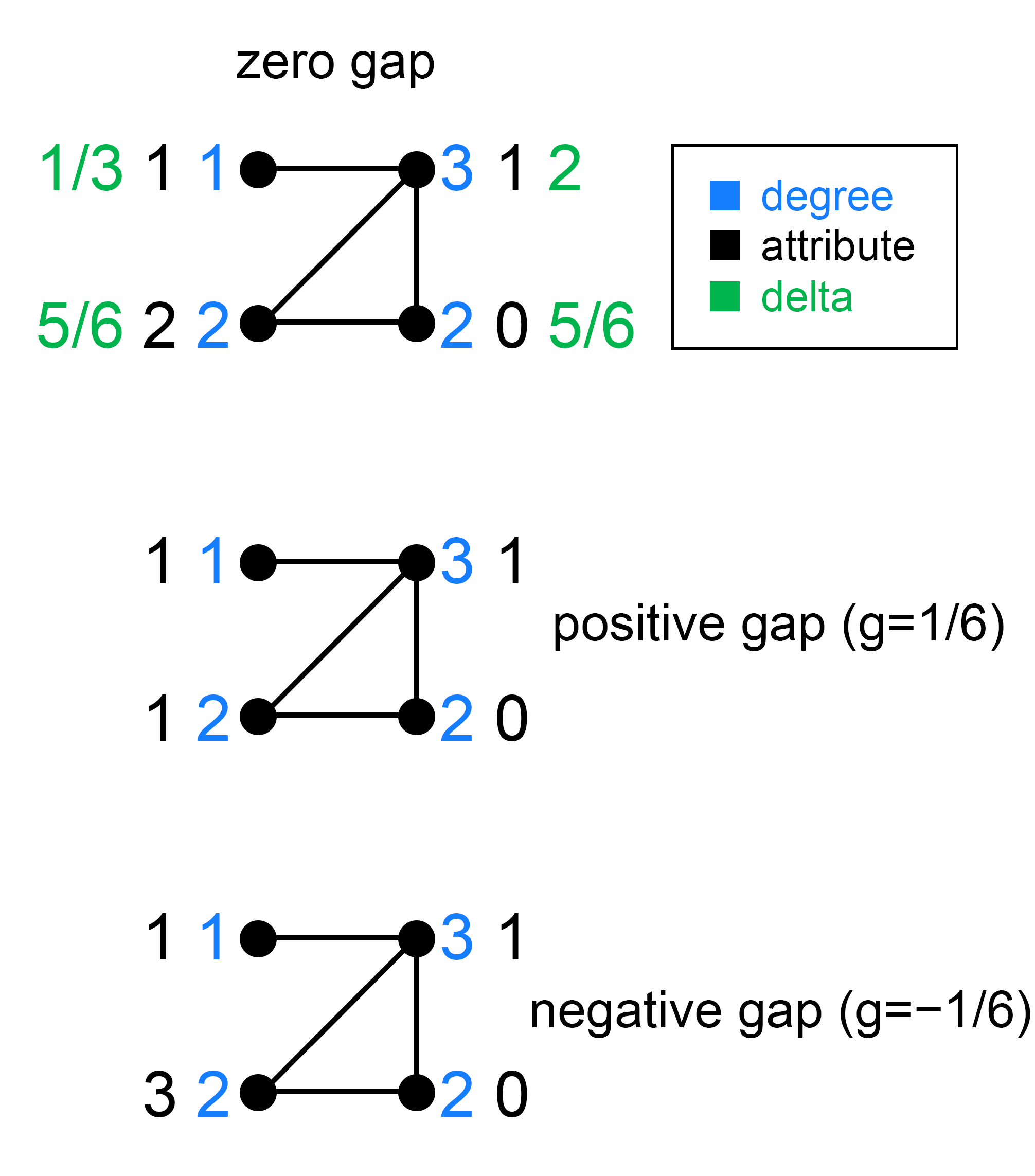}
\caption{
\label{fig:anti_sgfp}
An anti-SGFP graph such as this may produce differently-signed gaps for $r_{d,a}=0$, depending on the attribute sample.
}
\vspace*{-0.2in}
\end{figure}

As a final point, $r_{d,a}=0$ may, but wouldn't always produce 0 gaps for anti-SGFP topologies. For an example of that, see Figure~\ref{fig:anti_sgfp}. First note that this graph's $d_i$ and $\delta_i$ don't lie on a straight line, so $r_{d,\delta}\neq1$ and the graph is anti-SGFP. Then note that all 3 attribute samples (\{1,1,2,0\},\{1,1,1,0\},\{1,1,3,0\}) have correlation 0 with the degree sequence \{1,3,2,2\}. Finally, note the different signs of the gaps for the 3 cases. Here too, a 0 gap doesn't indicate regularity in degrees, or in attributes, or the graph being pro-SGFP. All pro-SGFP topologies would produce a gap of 0 if $r_{d,a}=0$, but anti-SGFP topologies may or may not do that.

\section{Pro-SGFP topologies aren't common}
\label{sec:pro_counting}
Given the Farkas' Lemma discussion in Section~\ref{sec:farkas}, if a graph has two differing degrees and two differing $\delta_i$ (true for all non-regular graphs), being pro-SGFP means having $r_{d,\delta}=1$. How common are pro-SGFP topologies in practice? To come up with a simple answer to that question, we use a $G_{n,p}$ random graph generator with $p=\frac{1}{2}$ which makes all edge arrangements (not to be confused with topologies, which are nodeID-agnostic) equally likely. We restrict the generator to only non-regular connected graphs and create 100,000 graphs for $n=3,4...10$. In addition to seeing what proportion of the graphs is pro-SGFP, we also want to find how high an $r_{d,a}$ an SGFP-failing attribute sample $a$ may have.

To see this, we use Simplex optimization (specifically the high performance dual revised simplex implementation algorithm) on the 4 constraints that we obtained in Section~\ref{sec:farkas}:
\begin{equation*}
\begin{cases}
\begin{aligned}
\sum_i a_i \geq 0 \\
\sum_i -a_i \geq 0 \\
\sum_i d_i a_i \geq 0 \\
\sum_i \delta_i a_i < 0

\end{aligned}
\end{cases}
\end{equation*}
The third constraint, given the first two, is the numerator of $r_{d,a}$. The denominator of $r_{d,a}$ is 1 if the sum of squares of the elements of $a$ is 1, or $\sum_i a_i^2 = 1$. Note that we can rescale the elements of any attribute sample so that $\sum_i a_i^2 = 1$, which doesn't change $r_{d,a}$ or the sign of the gap (see Supplementary Information for a proof). Hence we can assume our attribute sample $a$ follows $\sum_i a_i^2 = 1$ and $r_{d,a} = \sum_i d_i a_i$ (the third constraint). We'd like to maximize $\sum_i d_i a_i$ given constraints 1, 2, and 4.

The fourth constraint represents a negative gap. We change it to $\sum_i \delta_i a_i \leq -0.001$ for the purposes of linear optimization. Removing the strict inequality will not allow us to achieve a true maximum $r_{d,a}$ for which SGFP fails, but the optimization is still useful because the true maximum correlation is at least as large as what we find. Note also that there may not be an achievable maximum but instead a supremum, which is connected to our discussion of $\varrho_G$ in Section~\ref{sec:path_infinite}.

We call the correlation we obtain a \say{high correlation} instead of a \say{max correlation} and denote it $r^{(\textrm{high})}_G$ for a given graph $G$.

Table~\ref{table} provides the resulting proportions of pro-SGFP graphs as well as $r^{(\textrm{high})}_G$ and $r_{d,\delta}$ for pro- and anti-SGFP graphs. It's clear that as the number of nodes $n$ increases, pro-SGFP topologies are seen less and less often. As predicted, all the pro-SGFP topologies we found have $r_{d,\delta}=1$. Interestingly, $r_{d,\delta}$ is high even for anti-SGFP topologies. Naturally, for pro-SGFP topologies, $r^{(\textrm{high})}_G$ is not positive since no attribute sample with $r_{d,a}>0$ can fail SGFP for a pro-SGFP topology. For anti-SGFP topologies, $r^{(\textrm{high})}_G$ is around 0.25. 

While the pro-SGFP proportion goes down as $n$ grows, there are some non-trivial pro-SGFP topologies that can be easily described, such as \say{star} graphs and $K_n - e$ graphs as mentioned at the end of Section~\ref{sec:farkas}. Overall, though, it's hard to characterize the specific \say{look} of pro-SGFP graphs, so looking at whether $r_{d,\delta}$ is 1 or not is the way to go.

For $n=3$, the graph generator found only pro-SGFP topologies and no anti-SGFP topologies because the only 3-node connected non-regular topology is the 3-node path, which is $K_3 - e$.

For $n=4$, the pro-SGFP topologies we found are a path graph $w-x-y-z$; a star graph with $w$ connected to $\{x,y,z\}$ only; and a complete graph on 4 nodes with one edge removed (which is $K_4 - e$).

\begin{table}[h]
\label{table}
\begin{center}
\begin{tabular}{ | p{0.5cm} | p{2.5cm} | p{2.5cm} | p{2.5cm} | p{2cm} | p{2cm} | } 
 \hline
 \small $n$ & 
 \small proportion of pro-SGFP topologies (non-unique) & 
 \small mean $r^{(\textrm{high})}_G$ (pro-SGFP) & 
 \small mean $r^{(\textrm{high})}_G$ (anti-SGFP) & 
 \small mean $r_{d,\delta}$ (pro-SGFP) & 
 \small mean $r_{d,\delta}$ (anti-SGFP) \\ 
 \hline
 \normalsize
 3 & 1 & -0.0005 & N/A & 1 & N/A \\ 
 \hline
 4 & 0.6458 & -0.0005 & 0.2621 & 1 & 0.9622 \\ 
 \hline
 5 & 0.0900 & -0.0007 & 0.2265 & 1 & 0.9505 \\ 
 \hline
 6 & 0.0620 & -0.0006 & 0.2588 & 1 & 0.9398 \\ 
 \hline
 7 & 0.0023 & -0.0006 & 0.2584 & 1 & 0.9396 \\ 
 \hline
 8 & 0.0014 & -0.0006 & 0.2579 & 1 & 0.9402 \\ 
 \hline
 9 & 0.0001 & -0.0006 & 0.2519 & 1 & 0.9432 \\ 
 \hline
 10 & 2e-05 & -0.0010 & 0.2441 & 1 & 0.9471 \\ 
 \hline
 
\end{tabular}
\end{center}
\caption{\label{table} For $n=3,4..10$, we used the $G_{n,\frac{1}{2}}$ graph generator to find 100,000 instances of non-regular connected networks for each $n$. The table describes what proportion of those is pro-SGFP for each $n$; the correlations that may fail SGFP for pro- and anti-SGFP graphs for each $n$; and the degree-delta correlation $r_{d,\delta}$ for the two groups.}
\end{table}

\section{Failing SGFP with real network data}
\label{sec:real}
It's interesting whether attribute samples available as part of real-world data may fail SGFP. To provide an instance where that happens, we looked at 100 anonymized networks from the Facebook100 dataset
(see \cite{traud2012social,altenburger2018monophily} for details on this data). This is a useful dataset because with 100 disconnected school networks it provides multiple related instances. The average network size is 12,803.16.

\begin{figure}[t]
\centering
\includegraphics[width=0.4\textwidth]{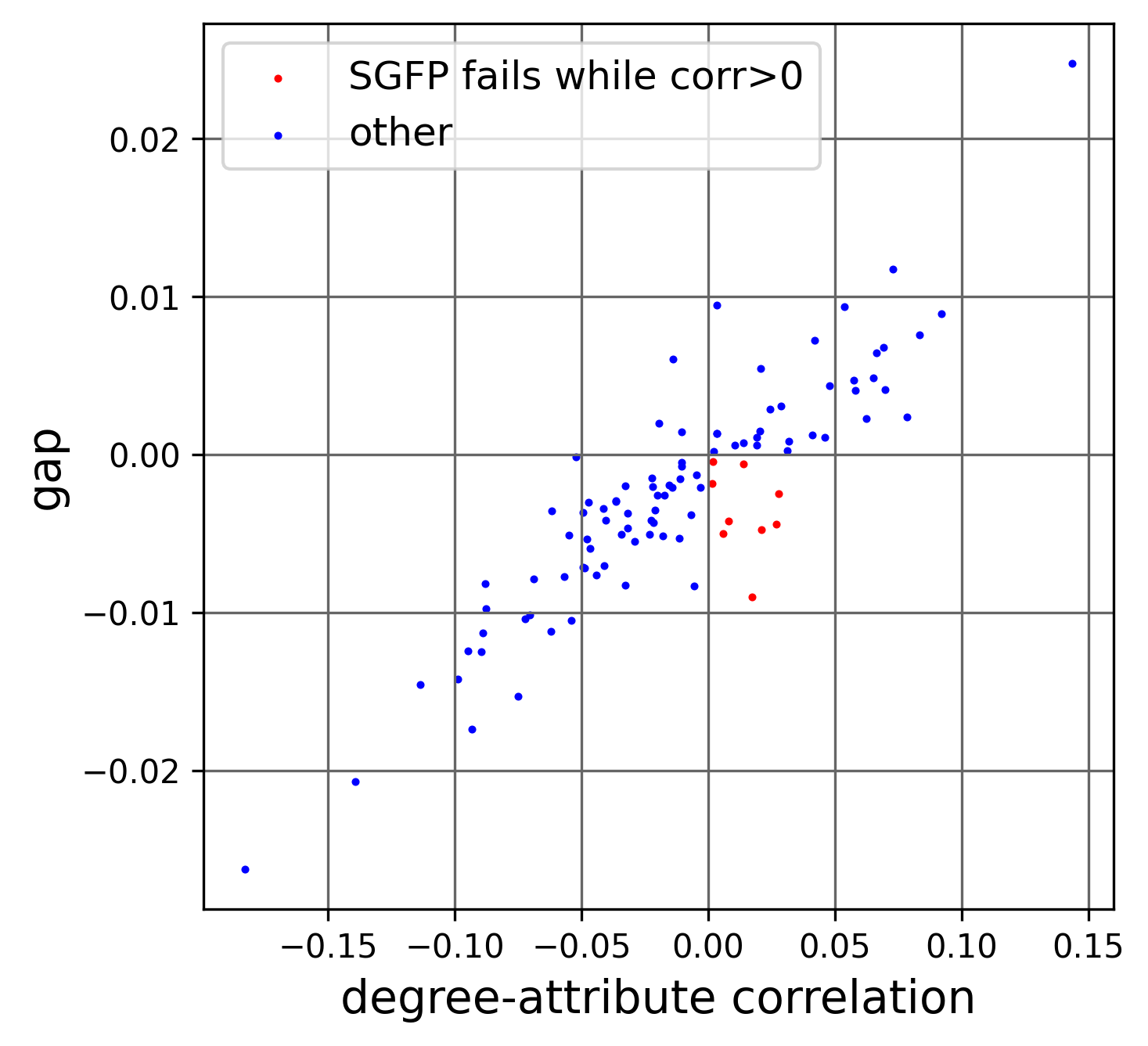}
\caption{
\label{fig:fb100}
Plotting $r_{d,a}$ and the gap for 100 Facebook100 networks with ``proportion of friends with the same reported gender'' (including NULL) as attribute. For 9 schools, the gap is negative while $r_{d,a}$ is positive. These are real-world examples of failing SGFP with attributes positively correlated with degree.}
\end{figure}

The dataset doesn't come with a lot of numerical attributes, but one useful field is gender, equal to \say{male,} \say{female} or \say{not reported}. We looked at one network at a time and, for each node $i$, created a data-driven attribute \say{prop\_own} which indicated what proportion of $i$'s friends shared its gender value. With the attribute sample in place, we were able to calculate the degree-attribute correlation $r_{d,a}$ and the SGFP gap. Figure~\ref{fig:fb100} plots those for the 100 schools. The 9 points in red represent schools with both positive $r_{d,a}$ and negative gap, demonstrating a failure of SGFP directly on these instances.

Note that we used data-driven attributes here. While only 9\% of Facebook100 networks fail SGFP in this case, all of these networks are anti-SGFP. This means that we can \textit{construct} SGFP-failing attribute samples with high $r_{d,a}$ for all these networks, which we do in the next section.

\section{Finding high SGFP-failing correlations for real networks}
\label{sec:real_optimize}
\begin{figure}[t]
\centering
\includegraphics[width=0.4\textwidth]{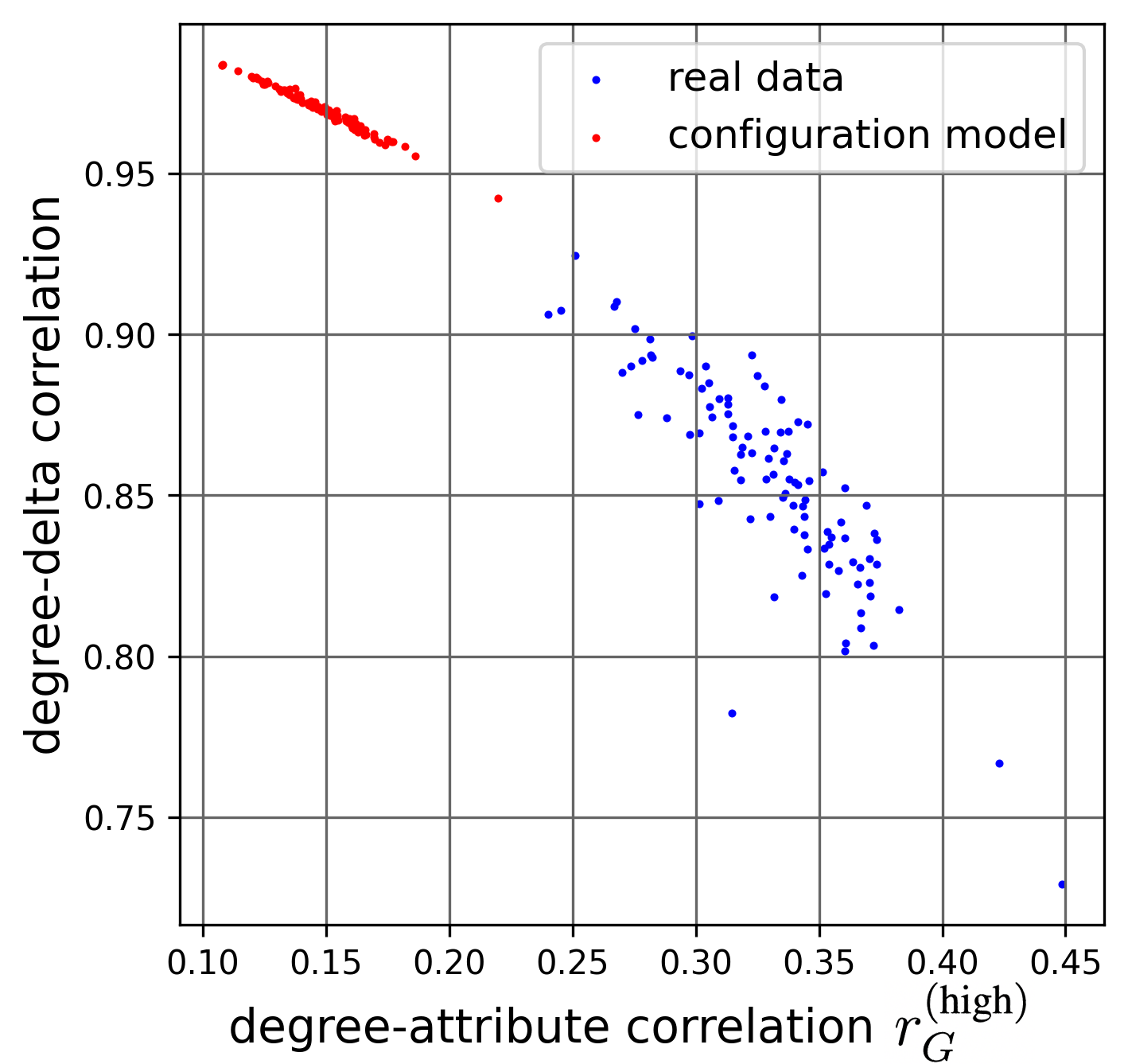}
\caption{
\label{fig:max_corr}
For Facebook100 data, it's possible to find SGFP-failing attribute samples with $r_{d,a}$ up to 0.45. When we remove social structure with a configuration model, the correlations we find with Simplex optimization are much lower.}
\vspace*{-0.2in}
\end{figure}

Another interesting question to ask of the Facebook100 data is how high $r_{d,a}$ can be for some SGFP-failing attribute sample $a$ assigned to a real graph that represents a school. To see this, we use the optimization technique introduced in Section~\ref{sec:pro_counting} and find $r^{(\textrm{high})}_G$ for each school. We expect a negative correlation between the $r_{d,\delta}$ and $r^{(\textrm{high})}_G$, since $r_{d,\delta}$ of 1 implies that a graph is pro-SGFP and its SGFP-failing correlation is bound by 0 from above (so $r^{(\textrm{high})}_G$ too is negative), but if $r_{d,\delta}$ is lower than 1, the graph is anti-SGFP and we could expect to find a positive SGFP-failing correlation through optimization (so $r^{(\textrm{high})}_G>0$). We find that $r^{(\textrm{high})}_G>0$ and $r_{d,\delta}<1$ for all networks, meaning they all are anti-SGFP. The blue points in Figure~\ref{fig:max_corr} represent $r^{(\textrm{high})}_G$ and $r_{d,\delta}$ for each school. The correlation for the blue points is -0.85, a strong linear relationship. 

It is also interesting whether this relationship is influenced by the social structure inherent in Facebook100 data. To check this, we rewire each network using the configuration model (removing any parallel edges and self-loops).
The resulting networks, like the original Facebook100 graphs, may contain isolates, but when computing the gap we can disregard them, and they also don't have weight in the optimization. We plot $r^{(\textrm{high})}_G$ vs $r_{d,\delta}$ for the rewired networks in red in Figure~\ref{fig:max_corr}. The correlation is -0.98 for the red points, and the $r_{d,\delta}$ values are similar to those of the random networks from Section~\ref{sec:pro_counting} (Table~\ref{table}). The striking difference between the two sets of points suggests that, given a degree sequence, social structure plays a role when it comes to higher possible SGFP-failing degree-attribute correlations. The results also suggest that, given a degree sequence, having social structure 
provides for smaller $r_{d,\delta}$.

\section{Conclusion}
In contrast to the Friendship Paradox and the List (network-level) Generalized Friendship Paradox, the Singular (node-level) Generalized Friendship Paradox is not a phenomenon that applies to networks in general, or regardless of their structure. The degree-attribute correlation is not the only parameter that determines whether SGFP holds, and the correlation being positive or very close to 1 doesn't mean SGFP won't fail.

If the network structure is pro-SGFP, SGFP will fail if and only if the degree-attribute correlation is negative. If the network structure is anti-SGFP, SGFP may fail for both negative and positive degree-attribute correlations; it would also hold for \textit{negative} correlations for attribute samples $-a$ where $r_{d,a}$ is positive and SGFP fails ($r_{d,-a}=-r_{d,a}$ and $g(-a)=-g(a)$). Given our data analysis and simulation, we conclude that anti-SGFP topologies are very common. How high SGFP-failing correlations can be depends on the specific anti-SGFP graph topology as shown in the discussion of $r^{(\textrm{high})}_G$ in Sections~\ref{sec:pro_counting} and \ref{sec:real_optimize}.

Taking all this together, we conclude that we can't simply assume that \say{your friends' attributes are greater than yours.}
Each real-world network we want to study needs to be checked for whether SGFP applies to it given its structure and attribute sample.

\section*{Data availability} 

The dataset we've used, Facebook100, is introduced in a paper by Traud et al. and is available from the authors of that work: \href{http://sciencedirect.com/science/article/abs/pii/S0378437111009186}{sciencedirect.com/science/article/abs/pii/S0378437111009186}

\section*{Supplementary information}
The supplementary information is available via \textit{Scientific Reports} at\\ \underline{\href{https://www.nature.com/articles/s41598-023-29268-7}{nature.com/articles/s41598-023-29268-7}}.

\end{document}